\documentclass[a4,aps,amsmath]{revtex4}
\usepackage{graphicx}
\usepackage{color}
\newcommand{\be}{\begin{equation}}
\newcommand{\ee}{\end{equation}}
\newcommand{\ba}{\begin{array}}
\newcommand{\ea}{\end{array}}
\newcommand{\bea}{\begin{eqnarray}}
\newcommand{\eea}{\end{eqnarray}}
\newcommand{\bdm}{\begin{displaymath}}
\newcommand{\edm}{\end{displaymath}}

\begin{document}

\title{Analysis of wasp-waisted hysteresis loops in magnetic rocks}

\author{R S Kharwanlang}
\affiliation{%
Physics Department \\ North Eastern Hill University \\ 
Shillong-793 022, India}%
\author{Prabodh Shukla}
\email{shukla@nehu.ac.in}
\affiliation{%
Physics Department \\ North Eastern Hill University \\ 
Shillong-793 022, India}%


\begin{abstract}

The random-field Ising model of hysteresis is generalized to dilute 
magnets and solved on a Bethe lattice. Exact expressions for the major 
and minor hysteresis loops are obtained. In the strongly dilute limit 
the model provides a simple and useful understanding of the shapes of 
hysteresis loops in magnetic rock samples.

\end{abstract}

\maketitle

\section{Introduction}

Hysteresis is a non-equilibrium effect~\cite{bertotti}. If a system is 
driven by a cyclic force that changes faster than the system can adjust 
to it, then the response does not move up and down on a single path but 
rather makes a hysteresis loop. Theoretically the area of the hysteresis 
loop should vanish as the frequency of the driving field goes to zero 
but several systems with quenched disorder show large hysteresis even at 
the slowest driving rate. This behavior arises from the presence of 
numerous metastable states in the system that are separated from each 
other by energy barriers much larger than the thermal energy. The 
metastable states correspond to local minima in the free-energy 
landscape of the system. The system remains practically trapped in a 
local minimum and is unable to attain thermal equilibrium over 
observation times. However, it can be made to jump from one local 
minimum to another if a sufficiently strong force is applied to it. We 
focus on magnetic systems. The magnetization induced by a cyclic field 
traces a hysteresis loop. The loop is essentially a locus of 
magnetizations of metastable states along the trajectory. Its shape and 
area are clearly objects of practical interest because these determine 
the rate of energy dissipation in the system. Somewhat less obvious but 
equally important is the fact that the hysteresis loop also contains 
information regarding the distribution of local free-energy minima and 
the energy barriers between them. Sethna et al introduced the 
non-equilibrium random-field Ising model of hysteresis that not only 
reproduces the shapes of hysteresis loops 'pleasantly familiar to the 
experimentalist' but also provides an understanding of other aspects 
associated with it e.g. Barkhausen noise, return point memory, and 
critical-point phenomena~\cite{sethna1,sethna2}.

Originally the random-field Ising model was introduced to study the 
effect of quenched positional disorder on the critical behavior of a 
system in thermal equilibrium ~\cite{imry-ma}. It showed that even an 
arbitrarily small amount of disorder raises the lower critical dimension 
of a system. The lower critical dimension is the dimension below which a 
system can not possess long-range order in a state of thermal 
equilibrium. Above its lower critical dimension, it may evolve into an 
ordered state at low temperatures but its approach to thermal 
equilibrium is not smooth. This is due to the presence of a large number 
of local minima in the free-energy landscape of disordered systems 
~\cite{young}. The local minima are surrounded by high barriers. The 
barrier heights are random but much higher than the thermal energy of 
the system. Hence the approach to thermal equilibrium is very slow and 
sporadic. Indeed the system may not reach equilibrium over practical 
time scales. This makes the determination of the equilibrium state (the 
global minimum of free-energy) a difficult task numerically as well as 
analytically. The dynamics of the approach to thermal equilibrium in the 
random-field Ising model has been well studied ~\cite{villain,dsfisher} 
but the progress has been slow due to the difficulty of the problem.

The difficulty of thermal equilibration is sidelined if the model is 
used to study hysteresis. Practically speaking, even in the zero 
frequency limit of the driving field, hysteresis is observed over time 
periods that are much shorter than the time required for the system to 
equilibrate. Thus the thermal relaxation process and the global minimum 
of the free-energy are not of primary importance in this case. In the 
non-equilibrium random-field Ising model proposed by Sethna et al 
~\cite{sethna1}, the time required by the system to equilibrate is set 
equal to infinity. Equivalently, the temperature of the system is set 
equal to zero. The metastable states under the stochastic thermal 
dynamics thus become stable states under the athermal zero-temperature 
deterministic dynamics. This does not compromise with the essential 
physics of the problem. There is an argument based on the 
renormalization group theory that the phase transition in the 
equilibrium random-field Ising model is controlled by a stable 
zero-temperature fixed point~\cite{dsfisher}. Several key features of 
hysteresis observed in disordered ferromagnets as well other systems 
whose dynamics is characterized by avalanches are very well reproduced 
qualitatively and even quantitatively by the zero-temperature 
non-equilibrium random field Ising model~\cite{sethna3}. The 
deterministic dynamics also makes the model amenable to an exact 
solution in some special cases~\cite{dhar,shukla,shukla2}.

This paper generalizes the non-equilibrium random-field model of 
hysteresis to dilute magnetic systems. We solve the dilute version of 
the model exactly on a Bethe lattice and apply the results to explain 
the shapes of hysteresis loops of magnetic rocks. We choose magnetic 
rocks for two reasons: (i) these are natural realizations of very dilute 
magnetic systems, and (ii) have not received the same attention in the 
physics literature as in geology. It is recognized that rock magnetism 
arises from a few percent or less of magnetic minerals present in the 
rocks~\cite{dunlop1,dunlop2}. The commonly occurring magnetic minerals 
are magnetite ($Fe_3O_4$), maghemite ($\gamma Fe_2O_3$), titanomagnetite 
($Fe_{2-y}Ti_yO_3$), pyrrhotite ($Fe_7S_8$), greigite ($Fe_3S_4$), 
hematite ($\alpha Fe_2O_3$), and goethite ($FeOOH$). The 
composition~\cite{note} is generally determined by breaking the rock 
sample. Recently, hysteresis measurements have been employed as a a 
possible nondestructive alternative. The motivation for this comes from 
the fact that the hysteresis loop of a rock sample as a whole is 
generally different from that of its magnetic constituents in their pure 
form. Pure magnetic minerals are mostly ferromagnetic or ferrimagnetic 
and therefore their hysteresis loops are similar to those of iron 
($Fe$). On the other hand, hysteresis loop of a rock sample as a whole 
can have unusual shapes including a wasp-waisted shape that is 
constricted in the middle~\cite{roberts1,roberts2,tauxe1}. The 
wasp-waisted shape is thought to arise from the fact that the grains 
dispersed in the rock have a distribution of sizes, domain states, and 
coercive fields~\cite{roberts1, roberts2, tauxe1, wasilewski, becker, 
pick1, pick2, bean, hejda, bennett, davies,tauxe2,kletetschka}. One 
would like to determine the distribution of magnetic grains from the 
hysteresis loop of the rock sample. It is not immediately obvious if 
this is feasible. A given distribution of grains and a set of 
interactions between them will produce a unique hysteresis loop. 
However, the inverse relationship of the hysteresis loop to the contents 
of the rock is not necessarily unique. This is because the detailed 
information concerning the grains has to be integrated over in order to 
obtain the hysteresis loop. Nevertheless there are practical advantages 
in exploring the extent to which we can deduce the composition of a rock 
from its hysteresis loop. As a first step towards this goal, one would 
like to study simple models of hysteresis and explore how their 
predictions vary with the parameters of the model.

There are a number of models of hysteresis in magnetic rocks in the 
field of geology. The majority of these use qualitatively similar 
assumptions and differ from each other only in detail. As an 
illustration we consider the model studied in reference ~\cite{tauxe1}. 
Grains of magnetic minerals of various sizes are assumed to be randomly 
dispersed in the rock. The grains are frozen in their positions but 
their magnetization can rotate or flip under a driving field as well as 
under thermal fluctuations. Interactions between the grains are 
neglected. This is presumably justified because the grains occupy only a 
few percent of the total volume of the sample and therefore they may be 
well separated from each other to interact significantly. The size of a 
grain determines the quality of its response to the applied field. Small 
grains (say 30 nm or less) behave like a paramagnet and respond to a 
cyclic field without hysteresis. Larger grains behave like a ferromagnet 
and respond to the cyclic field with hysteresis. This is because the 
size of the grain is related to the size of the energy barriers that 
stand in the way of smooth rotation or flipping of magnetization within 
a grain. These barriers are small for small grains and large for large 
grains. The thermal energy of the system provides the criterion for 
deciding whether a grain is small or large. If the barriers are small in 
comparison with the thermal energy, the magnetization is able to rotate 
freely under thermal fluctuations and is able to attain thermal 
equilibrium. In this situation the average magnetization is zero in zero 
applied field. In a non-zero applied field it is given by the Langevin 
function for a paramagnet. Larger grains are not able to attain thermal 
equilibrium over time scales of the experiment. Their response to the 
cyclic field is the non-equilibrium response in the form of a hysteresis 
loop. A hysteresis loop may be characterized by a number of parameters 
such as the saturation magnetization, the remanent magnetization, and 
the coercive field. These parameters are related to the size, shape and 
material of the grain. Some variants of the model of hysteresis in a 
rock sample consider directly a distribution of the coercivities of the 
grains, others a distribution of the sizes that is translated into 
coercivities through an assumed relationship between the two. The task 
of the models is to reproduce the experimentally observed shapes of 
hysteresis loops for a reasonable choice of the parameters of the model. 
Some models achieve this task by considering grains of different sizes 
~\cite{tauxe1,bean}, others by strongly contrasting coercivities 
~\cite{wasilewski,roberts1}. A few models have also included 
interactions between the grains ~\cite{bennett, hejda, davies}. A number 
of models reproduce major hysteresis loops similar to those observed in 
the laboratory experiments. Thus the difficulty is not that we do not 
have a model of rock magnetism but that we have several. The question 
naturally arises if we can distinguish between these models. Some 
authors have suggested that the comparison of experimental first-order 
reversal curves (minor hysteresis loops) with the predictions of 
different models may serve to distinguish between them~\cite{bennett}.

As stated earlier, we adapt the non-equilibrium random-field model of 
hysteresis to dilute magnets. It offers a framework for understanding 
hysteresis loops of magnetic rocks and the relationship between 
different models employed for the purpose in the field of geology. The 
non-equilibrium random field Ising model ~\cite{sethna1,sethna2} is 
defined on a lattice whose sites are occupied by an Ising spin (a binary 
variable that represents a unit domain with magnetization $\pm1$). The 
magnetization of a unit domain is allowed to flip (up/down) rather than 
rotate continuously. Each spin interacts with its nearest neighbors and 
experiences a uniform external field as well as a Gaussian quenched 
random-field with average value zero and standard deviation $\sigma$. 
This model along with the zero-temperature Glauber dynamics has played a 
key role in understanding several aspects of ferromagnetic hysteresis 
including Barkhausen noise, return point memory, and scale invariant 
avalanches characterizing critical hysteresis. The ingredient we add to 
this model is the random dilution of magnetic sites. We are interested 
in the limit of large dilution when only a few percent of the sites are 
occupied by spins. In this case the spins form small isolated clusters 
of different sizes. The absence of a large spanning cluster on the 
lattice precludes scale invariant avalanches or critical hysteresis but 
reproduces shapes of hysteresis loops that are commonly seen in magnetic 
rock samples. The size distribution of scattered clusters on the lattice 
is determined by the random occupancy of the lattice sites by spins. A 
cluster may be thought of as a magnetic grain in other models of rock 
magnetism but here there is no need to make a separate assumption for 
the distribution of grain size. It is also easily understood why small 
clusters behave somewhat like paramagnets and larger ones like 
ferromagnets. Spins flip up or down whenever the net field at their site 
changes sign. An isolated spin (smallest cluster) has no memory and 
behaves like a perfect paramagnet. A spin connected to other spins is 
influenced by them in addition to the on-site magnetic field and 
therefore it shows hysteresis. The random field Ising model of 
ferromagnetic hysteresis can be solved exactly on a Bethe lattice and 
gives important insights into the behavior of the model 
~\cite{dhar,shukla}. In the following we extend this solution to the 
dilute case. The exact solution in the dilute case is convenient in 
studying the effect of changing various parameters on the shape of 
hysteresis loops without performing time consuming simulations of the 
model.

\section{The model}

The dilute random-field Ising model is defined by the Hamiltonian,

\be H=-J\sum_{i,j}c_i c_j{S_i}{S_j} 
-\sum_{i}{h_i}c_i{S_i}-{h}\sum_{i}c_i{S_i}, \ee

where $J$ ($J>0$) is a ferromagnetic exchange interaction and the sum is 
over nearest neighbor sites $\{i,j\}$ of a lattice. The restriction to 
nearest neighbor interactions means that the model applies to systems in 
which long range dipolar interactions are negligible. $S_i=\pm 1$ is an 
Ising spin, $h_i$ is a random-field, and $h$ is a uniform external 
field. The random-field $h_i$ is drawn from a Gaussian distribution with 
mean value zero and standard deviation $\sigma$. The quantity $c_i$ is a 
random variable taking the value $1$ with probability $c$ and $0$ with 
probability $1-c$. Thus $c$ is the concentration of lattice sites 
occupied by spins. The quantities $\{h_i\}$ and $\{c_i\}$ are quenched 
and therefore remain unchanged under the evolution of the system. The 
spins are the dynamical variables. These are governed by the 
zero-temperature Glauber dynamics at discrete time steps $t$,

\be S_i(t+1)= \mbox{sign } \left[ J\sum_j c_j S_j(t) +h_i+h \right]\ee

The sum on the right-hand-side is over nearest neighbors of site $i$. 
At a fixed applied field $h$, this dynamics lowers the energy of the 
system iteratively and takes it to a stable fixed-point $\{S^*_i(h)\}$ 
such that for each lattice site $i$

\be S^*_i(h)= \mbox{sign } \left[ J\sum_j c_j S^*_j(h) +h_i+h 
\right]\ee

We are interested in the hysteresis loop when the applied field $h$ is 
cycled from $-\infty$ to $\infty$ and back to $-\infty$ infinitely 
slowly. Numerically, we start with a sufficiently large and negative 
$h$, so that the system is at the fixed-point $\{S^*_i(h=-\infty)=-1\}$. 
Now $h$ is increased slowly until this fixed point becomes unstable. We 
hold $h$ fixed at this value and allow the system to evolve under the 
iterative dynamics until it reaches a new fixed point. The new 
fixed-point is characterized by its magnetization per site $m^*(h)$,

\be m^*(h)= \frac{1}{N} \sum_i c_iS^*_i(h) \ee

The process is repeated i.e. we start with a fixed-point and 
increase $h$ until this fixed-point becomes unstable. We then hold $h$ 
fixed and allow the system to evolve to a new fixed-point 
characterized by a higher $m^*(h)$. Holding $h$ constant during the 
evolution of the system amounts to the assumption that the applied 
field varies infinitely slowly as compared with the internal 
relaxational processes of the system. The above process is repeated 
until all fixed-points in increasing applied field are determined. The 
trajectory of the magnetizations $m^*(h)$ of these fixed-points forms 
the lower half of the hysteresis loop. The upper half of the 
hysteresis loop is obtained similarly by starting with the fixed-point 
with $m^*=1$ and decreasing the field in smallest steps to obtain the 
sequence of fixed-points up to $m^*=-1$. If the size of the sample $N$ 
is sufficiently large so that finite size effects can be neglected, 
the magnetization $m^*_u(h)$ on the upper half of the hysteresis loop 
is related to the magnetization $m^*_l(h)$ on the lower half by the 
theoretical symmetry relation $m^*_u(h)=-m^*_l(-h)$. Hysteresis loops 
for the undiluted case ($c$=1) have been studied numerically on 
$d$-dimensional regular lattices, and by an exact solution on a Bethe 
lattice ~\cite{dhar,shukla}. In the following we extend the exact 
solution on the Bethe lattice to $c \le 1$ and apply it to rock 
magnetism.

\section{Hysteresis on a Bethe lattice}

A Bethe lattice is an infinite-size branching tree of coordination 
number $z$. It may be visualized as the deep interior of a large 
branching tree (Cayley tree) of the same coordination number. Figure (1) 
shows a small ( 4 levels) Cayley tree with $z=3$ drawn such that the 
lattice points at the bottom row (level 0) are at the surface of the 
tree, and the point at the top (level 3) is at the root (center) of the 
tree. All lattice points except on the surface have $z$ nearest 
neighbors. A lattice point on the surface has only one neighbor. 
Analysis of hysteresis on a Cayley tree is simpler because there are no 
closed loops on the lattice. However most of the lattice points lie on 
the surface and therefore special care has to be exercised to ensure 
that the results apply to the deep interior of the tree and are 
insensitive to conditions on the surface. We adopt two separate methods 
to eliminate the surface effects. In our analytic calculations we use 
recursion relations that take us from the surface towards the interior 
of the tree level by level. Fixed-points of these recursion relations 
are insensitive to the random-fields on the surface. In numerical 
simulations we employ a different strategy. We perform simulations on a 
surfaceless random graph of coordination number $z$. The two methods of 
eliminating surface effects are equivalent and therefore our theoretical 
results match our simulation results perfectly.

Consider a tree whose sites are randomly occupied by an Ising spin with 
probability $c$. If $c$ is slightly less than unity the lattice acquires 
holes (regions without spins). For values of $c$ closer to zero it 
breaks up into disjointed clusters of spins. It is not immediately 
obvious that the earlier method of recursion relations ~\cite{dhar} 
starting from the surface of a compact lattice ($c=1$) and moving 
towards its interior is still useful i.e. if it would yield fixed-points 
in spite of enhanced surface effects. But we find that the method works 
over the entire range $0 \le c \le 1$. We start with a diluted lattice 
and $h=-\infty$ so that all spins are down initially. The field is 
slowly ramped up from $-\infty$ to $h$ and we ask what fraction of spins 
are up at $h$. We choose a site at random in the deep interior of the 
tree and call it the central site. The probability that the central site 
is occupied by a spin is equal to $c$. We now calculate the probability 
that it is up as well. Each nearest neighbor of the central site if it 
is occupied by a spin forms the vertex of a subtree. The subtrees do not 
interact with each other except through the central site. Therefore the 
evolution on $z$ subtrees meeting at the central site is independent of 
each other as long as the central site does not flip up from its initial 
state. It is an important property of the model that the order in which 
the spins flip up does not influence the fixed-point at $h$ 
~\cite{dhar}, so we may assume that the central site is the last site to 
flip up at $h$ if it flips up at all. Before we can say whether the 
central sites may flip up or not, we need to know the conditional 
probability $cP^*(h)$ that a nearest neighbor of the central site is up 
given that the central site is down at applied field $h$.

$P^*(h)$ is the fixed point of the recursion equation,

\be
P^l(h)=\sum_{i=0}^{z-1}\binom{z-1}{i}{(1-c)}^i\left[ \sum_{j=0}
^{z-1-i}\binom{z-1-i}{j}\left\{c P^{l-1}(h)\right\}^j \left\{c - 
cP^{l-1}(h)\right\}^{z-1-i-j}  p_j(i;h) \right]
\ee

The above equation is written on the assumption that all spins in the 
system were down before being exposed to a field $h$. Spins on the 
surface had the first chance to flip up at $h$, then spins on level 1 
and so on. In this process of organized relaxation, spins up to level 
$l-1$ have been relaxed and we are at the point of calculating the 
probability that a spin at level $l$ (say at site $x$) flips up while 
spins at level $l+1$ are still down. With this explanation, equation (5) 
is easily understood after various symbols are defined. $P^l(h)$ is the 
conditional probability that the spin at site $x$ at level $l$ is up 
given that its nearest neighbor at level $l+1$ is down. Similarly, 
$P^{l-1}(h)$ is the conditional probability that a site at level $l-1$ 
is up given that its nearest neighbor at level $l$ (site $x$) is down. 
The site $x$ has one neighbor at level $l+1$ and $z-1$ neighbors at 
level $l-1$. These neighbors can be occupied by spins with probability 
$c$ or unoccupied with probability $1-c$. The neighbors at level $l-1$, 
if occupied, could be up or down independently of each other; $p_j(i;h)$ 
is the probability that site $x$ has sufficient quenched field to flip up 
at $h$ if $j$ neighbors are up, $i$ neighbors are unoccupied by spins, 
and consequently $z-j-i$ neighbors are down.

\be p_j(i;h)=\int_{(z-2j-i)J-h}^{\infty}\phi(h_i)dh_i; 
\hspace{.5cm} \phi (hi)= 
\frac{1}{\sqrt{2\pi\sigma^2}}e^{\frac{-hi^2}{2\sigma^2}} \ee

The probability that a randomly chosen site on the lattice (the 
central site) is occupied and up is,

\be p(h)=c\sum_{i=0}^{z}\binom{z}{i} \{1-c\}^i \left[ 
\sum_{j=0}^{z-i}\binom {z-i}{j} \{c P^*(h)\}^{j}\{c-cP^*(h)\}^{z-i-j} 
p_j(i,h) \right] \ee

The magnetization per spin on the lower and upper half of the major 
hysteresis loop is given by

\be m^*_l(h)=2p(h)-c; \mbox{\hspace{.5cm}}m^*_u(h)=-m^*_l(-h) \ee

If we reverse the applied field before completing the lower half of 
the major loop, we generate a minor hysteresis loop. First reversal of 
the field generates the upper half of the minor loop, and a second 
reversal generates the lower half. When the field on second reversal 
reaches the point where the first reversal was made, the lower half of 
the minor loop meets the starting point of the upper half. In other 
words, the minor loop closes upon itself at the point where it 
started. This property of the random-field Ising model is known as the 
return point memory. Consider the upper half of the minor loop. 
Suppose the applied field is reversed from $h$ to $h^d$ $(h^d \le h)$. 
We need to calculate the probability that an occupied site, say site 
$x$, that is up at $h$ turns down at $h^d$. When site $x$ turns up at 
$h$, the field on its nearest neighbors increases by an amount $2J$. 
This may cause some neighbors to turn up as well. Each neighbor that 
turns up increases the field on site $x$ by an amount $2J$. Therefore 
on reversing the field, site $x$ can turn down only after all 
neighbors which turned up after it have turned down. The probability 
$D^*(h^d)$ that an occupied nearest neighbor of site $x$ that was down 
before site $x$ turned up at $h$ is again down at $h^d$ is determined 
by the fixed point of the following recursion relation.

\be 
\begin{split} 
D^*(h^d)=c\sum_{i=0}^{z-1}\binom{z-1}{i}{(1-c)}^i 
\left[ \sum_{j=0}^{z-1-i} \binom{z-1-i}{j}\{c P^*(h)\}^j 
\{c-cP^*(h)\}^{z-1-j-i} \{1-p_{j+1}(i;h)\} \right] \\ 
+ c \sum_{i=0}^{z-1} 
\binom{z-1}{i}{(1-c)}^i \left[ \sum_{j=0}^{z-1-i} 
\binom{z-1-i}{j}\{c P^*(h)\}^j 
\{D^*(h^d)\}^{z-1-j-i}
\{ p_{j+1}(i;h)- p_{j+1}(i;h^d)\} 
\right]
\end{split} 
\ee

Given an occupied site $x$ that is up at $h$, the first sum above 
gives the conditional probability that an occupied nearest neighbor of 
$x$ remains down at $h$ after site $x$ has turned up. The second sum 
takes into account the situation that the nearest neighbor in question 
turns up at $h$ after site $x$ turns up but turns down at $h^d$.

The fraction of occupied sites that turn down at $h^d$ is given by

\be
q(h^d)=c \sum_{i=0}^{z}\binom{z}{i}{(1-c)}^i\left[ \sum_{j=0}^{z-i}
\binom{z-i}{j}\{c P^*(h)\}^j\{D^*(h^d)\}^{z-i-j}\{p_j(i,h) 
-p_j(i,h^d)\}\right]
\ee

The magnetization on the upper return loop and in the range of the 
applied field $h-2J \le h^d \le h$ is given by

\be
m(h^d)=2\{p(h)-q(h^d)\}-c
\ee

At $h^d=h-2J$, all occupied neighbors of the central site that flipped 
up because the central site flipped up at $h$ would flip down but the 
central site would stay up. If the applied field decreases further, 
the central site would turn down before any of its occupied nearest 
neighbors. This means that at $h-2J$ the system arrives at some point 
on the upper half of the major hysteresis loop, and moves on it upon 
further decrease in the applied field . The magnetization for $h^d \le 
h-2J$ is given by

\be m(h^d)=2\tilde{p}(h^d)-c \ee 

Where,
 
\be \begin{split} 
\tilde{p}(h^d)=c \sum_{i=0}^{z}\binom{z}{i}{(1-c)}^i \left[ 
\sum_{j=0}^{z-i}\binom{z-i}{j}\{c \tilde{P}^*(h^d)\}^j\{(1- 
c\tilde{P}^*(h^d))\}^ {z-i-j}p_j(i;h^d) \right] 
\end{split} \ee

and $\tilde{P}^*(h^d)$ is given by the fixed point of the following 
recursion relation.

\be
\begin{split}
\tilde{P}^l(h^d)=\sum_{i=0}^{z-1}\binom{z-1}{i}{(1-c)}^i \left[
\sum_{j=0}^{z-1-i}\binom{z-1-i}{j}\{c 
\tilde{P}^{l-1}(h^d)\}^j
\{c-c \tilde{P}^{l-1}(h^d)\}^{z-1-i-j} p_{j+1}(i;h^d) \right]
\end{split}
\ee

The lower half of the return loop is obtained by reversing the field 
from $h^d$′ to $h^u$ ($h^u > h^d$). If $h^d′< h-2J$, the lower half of 
the minor loop starts from the major loop, and therefore it is related 
by symmetry to the upper half of the return loop that has been 
obtained above. We only need to consider the case $h^d \ge h-2J$. In 
this case, the magnetization on the lower half of the return loop may 
be written as,
 
\be m(h^u)=2\{p(h)-q(h^d)+p'(h^u)\}-c \ee

Where $p'(h^u)$ is the probability that an occupied site that is up at 
$h$, down at $h^d$, turns up again at $h^u$.

\be
p'(h^u)=c \sum_{i=0}^{z}\binom{z}{i}{(1-c)}^i \left[ \sum_{j=0}^{z-i}
\binom{z-i}{j}\{U^*(h^u)\}^j\{D^*(h^d)\}^{z-i-j}
\{p_j(i,h^u)-p_j(i,h^d)\} \right]
\ee

Here $U^*(h^u)$ is the conditional probability that an occupied 
nearest neighbor of a site $x$ turns up before site $x$ turns up on 
the lower return loop. It is determined by the fixed point of the 
following equation.

\be
\begin{split}
U^*(h^u)=c P^*(h)- c \sum_{i=0}^{z-1}\binom{z-1}{i}{(1-c)}^i
\left[ \sum_{j=0}^{z-1-j}\binom{z-1-i}{j}\{cP^*(h)\}^j\{D^*(h^d)\}^
{z-1-i-j}\{p_j(i,h)-p_j(i,h^d)\} \right] \\
+ c 
\sum_{i=0}^{z-1}\binom{z-1}{i}{(1-c)}^i \left[ 
\sum_{j=0}^{z-1-i}\binom
{z-1-i}{j}\{U^*(h^u)\}^j\{D^*(h^d)\}^{z-1-i-j}
\{p_j(i,h^u)-p_j(i,h^d)\} \right]
\end{split}
\ee

The rationale for the equation (17) is as follows. Given an occupied 
site $x$ that is down at $h^d$, the first two terms on the right hand 
side account for the probability that an occupied nearest neighbor of 
site $x$ is up at $h^u \ge h^d$. Note that the neighbor in question 
must have been up at h in order to be up at $h^d$, and if it is 
already up at $h^d$, then it will remain up on the entire lower half 
of the return loop, i.e. at $h^u \ge h^d$. The last term gives the 
probability that the nearest neighbor was down at $h^d$, but turned up 
on the lower return loop before site $x$ turned up. It can be verified 
that the lower return loop meets the lower major loop at $h^u= h$ and 
merges with it for $h^u > h$, thus proving the property of return 
point memory.

The method of calculating the minor loop may be extended to obtain a 
series of smaller minor loops nested within the minor loop obtained 
above. The key point is that whenever the applied field is reversed, a 
site $x$ may flip only after all neighbors of site $x$ which flipped 
in response to the flipping of site $x$ on the immediately preceding 
sector have flipped back. The neighbors of site $x$ which did not flip 
on the preceding sector in response to the flipping of site $x$ do not 
flip in the reversed field before site $x$ has flipped. We have 
obtained above expression for the return loop when the applied field 
is reversed from $h$ on the lower major loop to $h^d$′ ($h-2J \le h^d′ 
\le h$), and reversed again from $h^d$′ to $h^u$′′ ($h^u′\le h$). When 
the applied field is reversed a third time from $h^u$ to $h^{dd}$ 
($h^{dd} < h^u$), expressions for the magnetization on the nested 
return loop follow the same structure as the one on the trajectory 
from $h$ to $h^d$.

The analytic results obtained above are depicted in Figure (2) for 
$\sigma=1.7$ and $z=4$ for two concentrations of magnetic minerals; (i) 
c=1 (no dilution), and (ii) c=0.8 . The fixed point of equation (5) is 
evaluated numerically for a sufficiently large and negative applied 
field $h$, and used in equations (7) and (8) to obtain the magnetization 
at the start of the lower half of the hysteresis loop 
($m_l^*\approx-c$). The applied field $h$ is then increased in small 
steps, and the process is repeated to determine $m_l^*(h)$ along the 
lower half of the hysteresis loop. Similarly, minor loops are obtained 
using equations (9) to (17). The upper half of the hysteresis loop is 
obtained by the relation $m_u^*(h)=-m_l^*(-h)$. Hysteresis on a $z \ge 
4$ lattice is qualitatively different from the case of $z=2$ and $z=3$. 
For $z \ge 4$ there exists a non-equilibrium critical point 
($\sigma=\sigma_c,h=h_c$) on the lower half of the hysteresis loop, and 
another symmetrically placed critical point on the upper half. At these 
non-equilibrium critical points the response of the system to a slowly 
varying driving field is singular. Critical points do not exist on 
lattices with $z=2, 3$. Let us focus on the critical point on the lower 
half of the hysteresis loop. For $z=4$ and $c=1$, we have 
$\sigma_c\approx1.78$. For $z > 4$, $\sigma_c$ increases with increasing 
$z$. For $\sigma < \sigma_c$, the two halves of the hysteresis loop have 
first order jumps in the magnetization as shown in Figure (2). For 
$c<1$, there is extra disorder in the system on account of the dilution 
of magnetic sites. This causes the first order jumps at $\sigma=1.78$ 
for $c=1$ to vanish at $c=0.8$, and the major hysteresis loop becomes 
smooth as seen in Figure (2). We have also shown two minor hysteresis 
loops for $c=0.8$. If $c<1$ but sufficiently large to form spanning 
clusters of occupied sites on the lattice, the qualitative behavior of 
hysteresis is similar to that on the undiluted lattice with $c=1$ but 
with an effectively reduced value of $\sigma_c$. These results are 
verified by numerical simulations of the model for the corresponding 
choices of the parameters $z, J, c$, and $\sigma$. As stated earlier, we 
performed numerical simulations on random graphs to eliminate surface 
effects. A random graph of N sites has no surface, but the price we pay 
is that it has some loops. However, for almost all sites in the graph, 
the local connectivity up to a distance of $log_{(z-1)}N$ is similar to 
the one in the deep interior of the branching tree. Thus the the 
theoretical results depicted in Figure (2) to (4) perfectly matched the 
corresponding results from the simulations of the model on random 
graphs. Indeed, the theoretical and simulation results are 
indistinguishable on the scale of the figures.

\section{Application to rock magnetism}

There are a variety of models in the literature that explain the shape 
of hysteresis loops in magnetic rock samples. Our objective is not 
merely to add to this list of models but also to simplify the situation. 
Our point is that extant models of ferromagnetic hysteresis also explain 
hysteresis in magnetic rocks if we add the ingredient of dilution to 
them. We have chosen the random field Ising model of hysteresis in 
ferromagnets at zero temperature and in the limit of zero frequency of 
the driving field. The advantage of this model is that it can be solved 
exactly on a Bethe lattice of an arbitrary coordination number 
$z$~\cite{shukla}. The model has only a few parameters; the coordination 
number $z$ of the lattice, the interaction energy $J$ that aligns 
nearest neighbor spins parallel to each other, and a parameter $\sigma$ 
that tends to disorder the system. In spite of its simplicity, this 
model explains a number of experimental observations regarding return 
point memory, Barkhausen noise, non-equilibrium critical points on 
hysteresis loops, and universal behavior in the vicinity of these 
points~\cite{sethna1,sethna2}. All we have done is diluted this model, 
i.e. only a fraction $c$ of the sites are occupied by magnetic entities. 
We have solved the dilute model exactly. Drawing upon the empirical 
observation that the magnetic minerals in a rock account for only a few 
percent of its mass, one may ask if the predicted hysteretic behavior of 
our model in the range $0.01 \le c \le 0.1$ is reasonably close to the 
observed behavior of magnetic rocks. It indeed appears to be the case 
considering the minimal number of free parameters in our model.

Figure (3) shows three hysteresis loops for $c=0.1$; and $\sigma=0.05, 
0.2$, and $0.5$ respectively. We have set $J=1$ for convenience. Let us 
focus on $\sigma=0.05$ first. As the applied field $h$ increases from 
$h=-\infty$ the magnetization stays at its saturation value $m=-0.1$ 
until $h\approx -3\sigma$. It rises sharply from $m \approx -0.1$ to $m 
\approx 0.03$ around $h \approx 0$, and from $m \approx 0.03$ to $m 
\approx 0.1$ around $h \approx 1$. The rise around $h=0$ is due to 
isolated spins in the system. The quenched field has a Gaussian 
distribution with average value zero and standard deviation $\sigma$. 
Therefore some of the isolated sites turn up at $h\approx -3\sigma$, and 
almost all of them are up at $h \approx 3\sigma$. The fraction of 
isolated sites is $c(1-c)^4$. This is approximately equal to $0.034$. 
Occupied sites in clusters of size larger than unity account for the 
remaining fraction $0.066$. After all isolated sites have turned up the 
magnetization per site increases to $-0.1+2\times 0.34=.032$. It stays 
at this plateau value until the applied field increases further by an 
amount $2J-3\sigma$ enabling some surface sites of a connected cluster 
of down spins (dangling bonds) to turn up. Nearly all sites turn up at 
$h=2J+3\sigma$ and we get the saturated magnetization $m=.1$ as shown in 
figure (3). The same mechanism applies to hysteresis loops for other 
values of $\sigma$. With increasing $\sigma$ the sharp corners tend to 
get more rounded and the middle (pinched) portion of the loop broadens. 
We get a gentle pinched loop for $\sigma=0.2$. For $\sigma=0.5$ the 
constriction in the middle disappears completely and we get a normal 
ferromagnetic type of hysteresis loop as shown in figure (3). Thus our 
model reproduces various shapes of hysteresis loops observed in magnetic 
rocks with a minimum number of parameters and without invoking different 
mechanisms for different shapes. Figure 4 (see caption) shows another 
way of representing the same data as shown in figure 3. In this 
representation, the constriction in the middle of the hysteresis loop 
appears as a more pronounced dip in the middle of the plot and therefore 
it may be of some value in the analysis of experimental data.

Can we deduce the magnetic composition of the rock from the shape of its 
hysteresis loop? This is not possible in general because the hysteresis 
loop contains the information in an integrated form. Information is 
irreversibly lost in the process of integration and can not be 
retrieved. There is no unique way to obtain the components of a sum from 
the sum! However, our model provides a few guiding principles. 
Ferromagnetic shapes of hysteresis loops indicate a larger random-field 
disorder $\sigma$ in the system. On the other hand a wasp-waisted loop 
especially with a long and narrow waist and sharp bends indicates 
relatively small disorder $\sigma$. In this case the connectivity of 
spins at the surface of clusters has the larger effect on the shape of 
hysteresis loops. It should be possible to deduce the magnitude of 
interaction $J$ from the locations of sharp turns in the magnetization 
along the applied field. Similarly, it should be possible to deduce the 
relative size of clusters from the plateaus in the magnetization. The 
distribution of cluster sizes and the plateaus of the magnetizations are 
known exactly in our model in terms of the two parameters ($J$ and 
$\sigma$) of the model. However, more complex rock samples may not be 
adequately characterized by a simple two-parameter Hamiltonian. Also, 
there may be systems whose frozen disorder is not represented adequately 
by a Gaussian distribution. We have focused on the Gaussian distribution 
because it applies to many systems ~\cite{sethna1,sethna2} and its use 
is justified by the central limit theorem. The effect of the shape of 
random-field distribution on critical hysteresis in the random-field 
Ising model ($c=1$) has been examined by Liu and Dahmen~\cite{liu}. They 
find that the Lorentzian and parabolic distributions of random fields 
yield the same critical exponents in three dimensions as the Gaussian 
random fields. This is not entirely surprising because a renormalization 
group analysis in $6-\epsilon$ dimensions ~\cite{dahmen} shows that for 
random-field distributions with a single maximum, it is only the 
curvature of the distribution at its maximum that determines the 
critical exponents. Therefore it is reasonable to use the Gaussian 
distribution as a common representative of smooth single-peaked 
distributions which all give the same critical exponents that agree with 
experiments. In the same vein, we have used the Gaussian distribution to 
understand the shapes of hysteresis loops in the strongly diluted limit 
of the random-field model. Before ending, we also wish to mention that 
wasp-waisted loops are not a property of magnetic rock samples only. 
Such loops are also seen in random magnets with the order parameter 
having a continuous symmetry~\cite{shukla2}, shape memory 
alloys~\cite{straka,murray}, and martensites~\cite{goicoechea}. Indeed 
the physics behind hysteresis in the random field Blume-Emery-Griffiths 
model~\cite{goicoechea} is very similar to that discussed here in the 
case of the dilute random-field Ising model.

\acknowledgments

We thank Deepak Dhar for reading an earlier version of this manuscript 
and making useful comments.

\begin{figure}[htb] 
\includegraphics[width=.9\textwidth,angle=0]{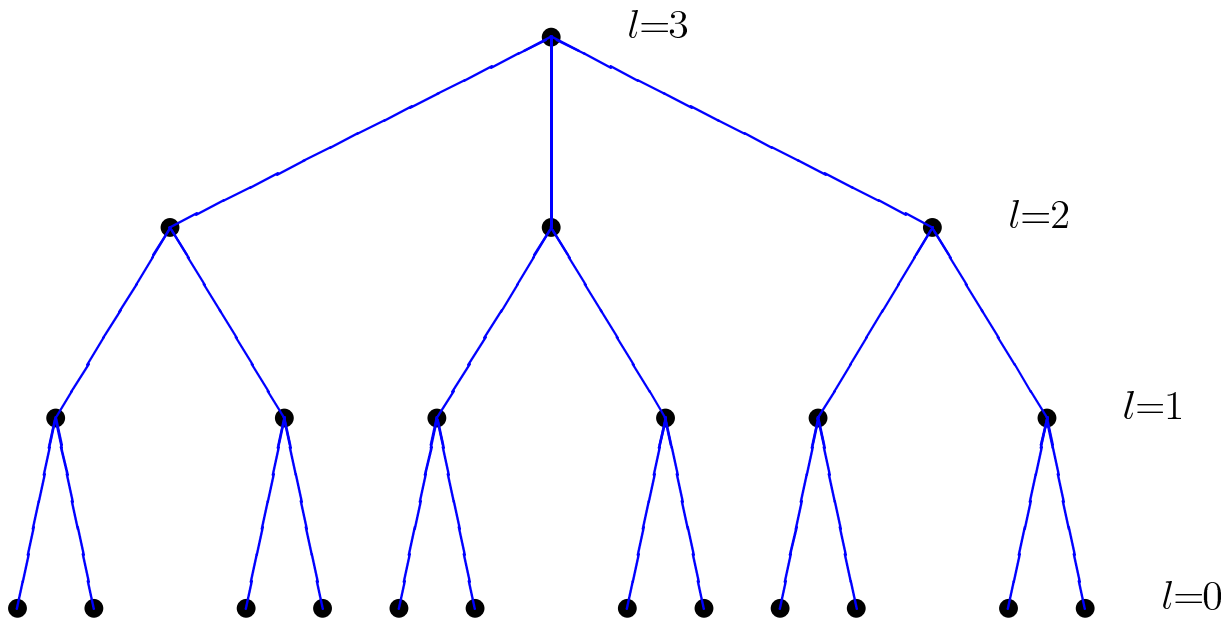}
\caption{(Color online) A Cayley tree with with 4 levels ($l=0,1,2,3$) 
and coordination number $z=3$. Each node is connected to $z$ nearest 
neighbors except the nodes at the surface ($l=0$) which have only one 
neighbor. The deep interior of the tree where surface effects can be 
neglected is known as the Bethe lattice.}
\label{fig1} 
\end{figure}

\begin{figure}[htb] 
\includegraphics[width=.9\textwidth,angle=0]{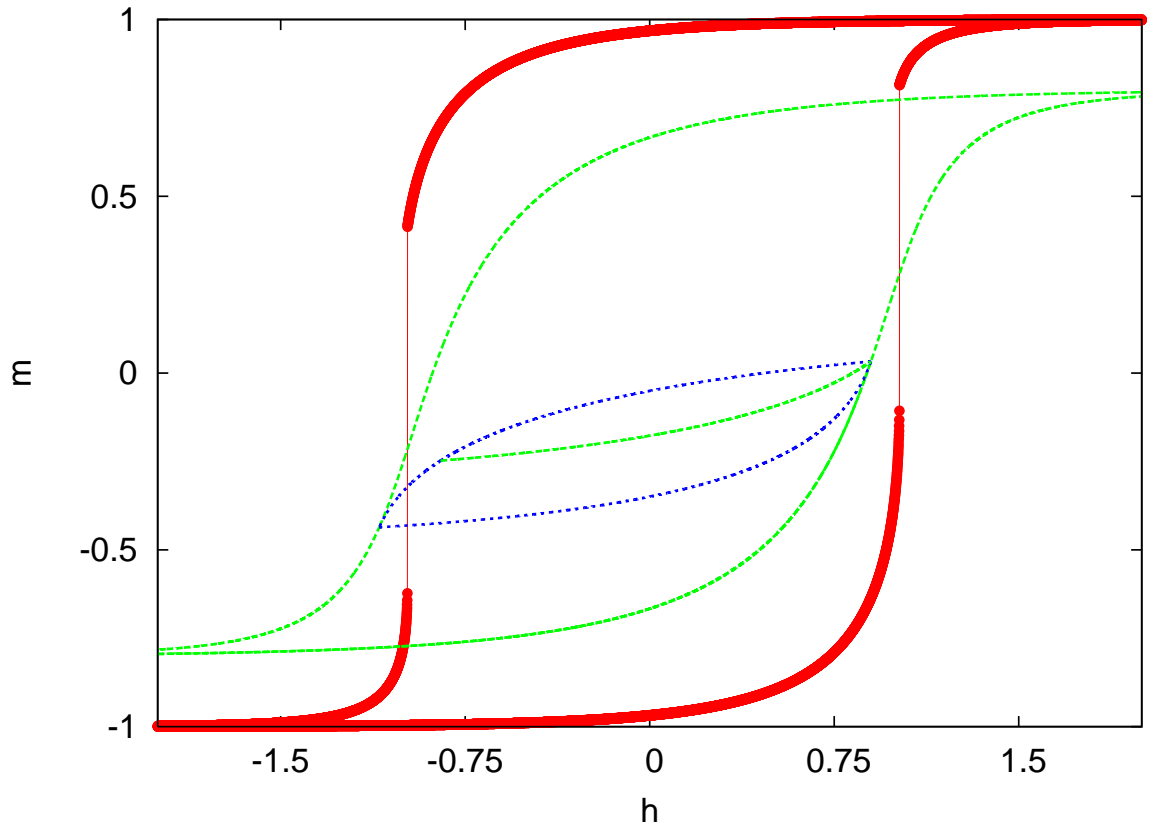}
\caption{(Color online) Hysteresis loops on a $z=4$ lattice for 
$\sigma=1.7$: The larger loop (red/dark) with saturation at $m=\pm1$ and 
first-order jumps in the magnetization is for $c=1$ (undiluted lattice); 
the smaller loop (green/grey) with saturation at $m=\pm0.8$ is for 
$c=0.8$; Two minor loops within the smaller loop are also shown which 
are obtained by making excursions from $h=0.9$ to $h=-1.1$ and $h=-0.85$ 
respectively.
}
\label{fig2} 
\end{figure}

\begin{figure}[htb] 
\includegraphics[width=.9\textwidth,angle=0]{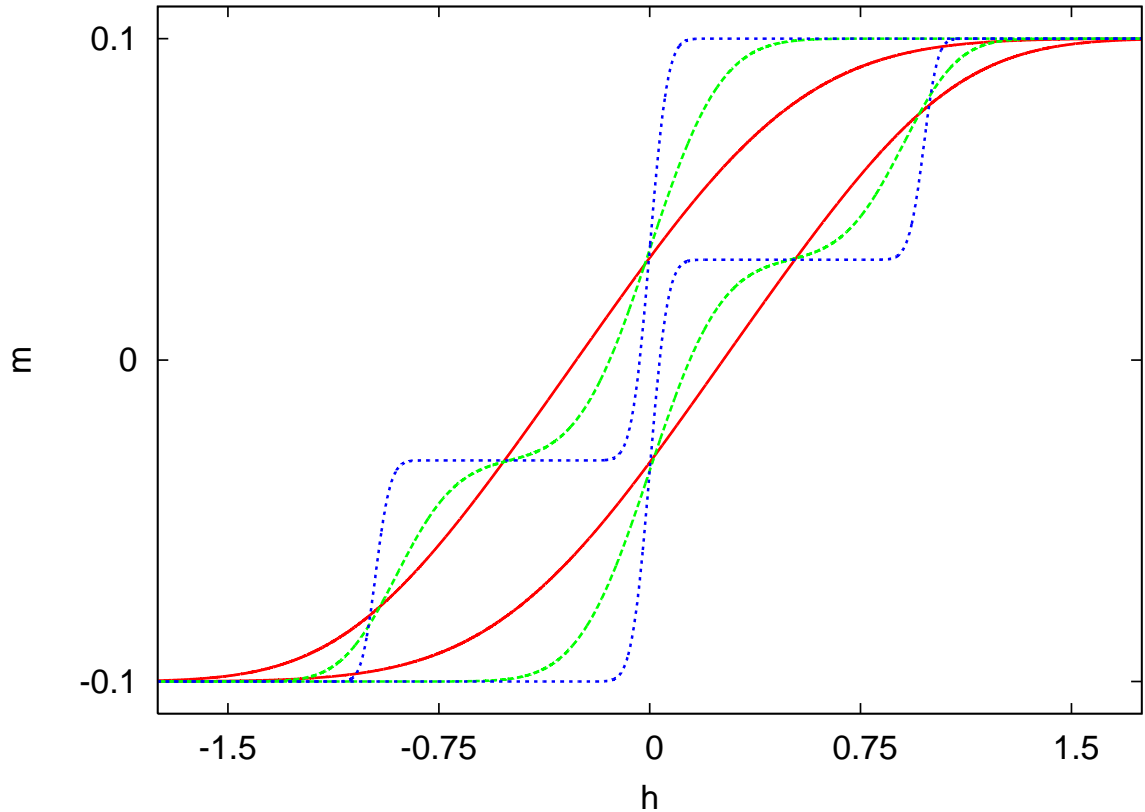}

\caption{(Color online) Hysteresis loops on a lattice with $z=4$ and 
$c=0.1$ as may be appropriate for magnetic rock samples. The saturation 
$m=\pm 0.1$ is controlled by the fractional content of magnetic minerals 
in the rock ($c=0.1$) but the shapes depend on $\sigma$. Three 
representative cases are shown: (i) ferromagnetic shape shown in red 
(loop with the largest width at $m=0$ for $\sigma=0.5$; (ii) a very 
pronounced wasp-waisted loop in blue (loop with the smallest width at 
$m=0$) for $\sigma=0.05$, and (iii) a wasp-waisted loop in green (having 
intermediate width at $m=0$ for $\sigma=0.2$).}
\label{fig3} 
\end{figure}

\begin{figure}[htb]
\includegraphics[width=.9\textwidth,angle=0]{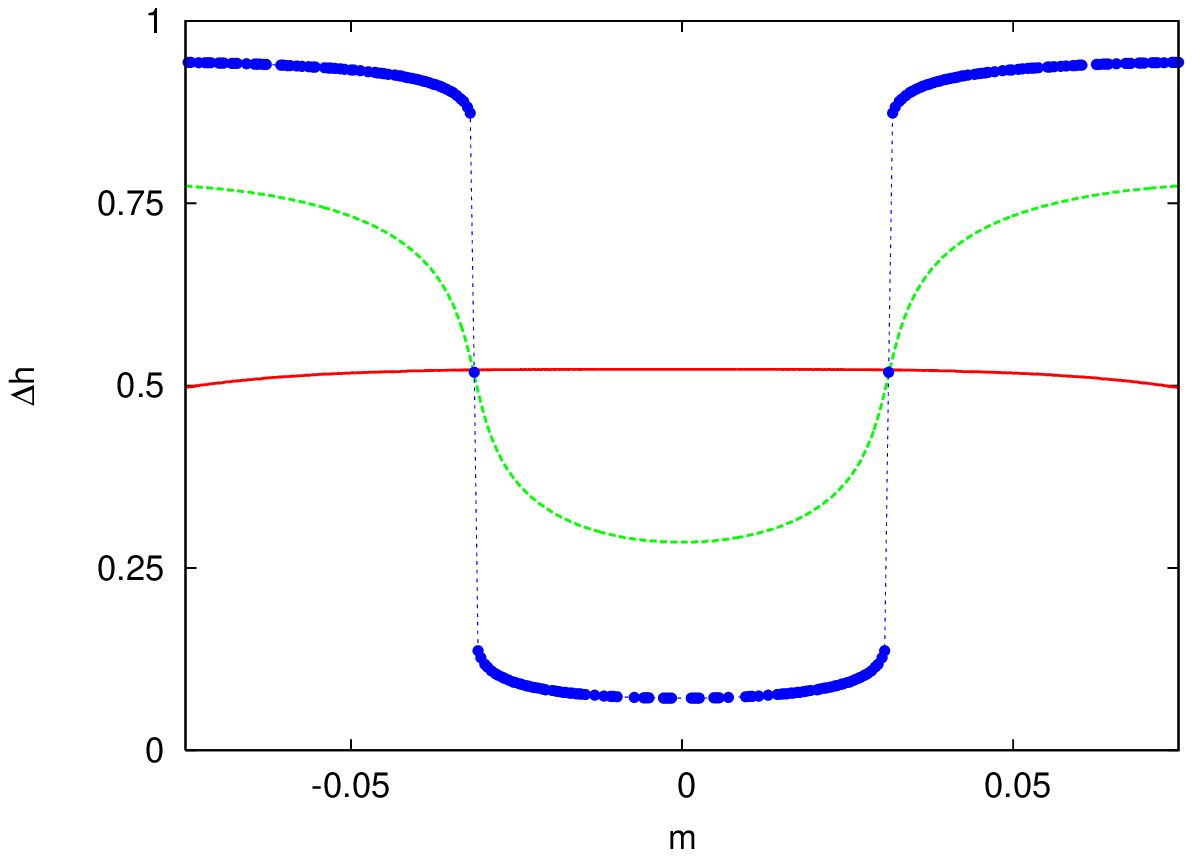}
\caption{(Color online) This is a different representation of figure 
(3). For a fixed magnetization $m$ along the $x$-axis, the $y$-axis 
shows the difference of corresponding applied fields on the upper and 
lower halves of the hysteresis loop. The red line (nearly horizontal 
thin line) corresponds to $\sigma=.5$, the blue line (thick line with 
two nearly vertical jumps) corresponds to $\sigma=0.05$. The green line 
(intermediate dotted line) corresponds to $\sigma=0.2$.}
\label{fig4}
\end{figure}


\begin{thebibliography}{99}

\bibitem{bertotti} See, for example, {\em{The Science of Hysteresis}}, 
edited by G Bertotti and I Mayergoyz, (Academic Press, Amsterdam, 2006).

\bibitem{sethna1} J P Sethna, K Dahmen, S Kartha, J A Krumhansl, B W 
Roberts, and J D Shore, Phys Rev Lett {\bf 70}, 3347 (1993).

\bibitem{sethna2} J P Sethna, K Dahmen, O Perkovic. \newblock The 
Science of Hysteresis (Elsevier Inc 2005) Vol 2. Ch 2. pp 107-179.

\bibitem{imry-ma} Y Imry and S -k Ma, Phys Rev Lett {\bf 35}, 1399 (1975).

\bibitem{young} D P Belanger and T Nattermann, in {\em{Spin Glasses and 
Random Fields}} edited by A P Young ( World Scientific, Singapore, 1998 ).

\bibitem{villain} J Villain, Phys Rev Lett 52, 1543 (1984); G Grinstein 
and J F Fernandez, Phys Rev B 29, 6389 (1984); A J Bray and M A Moore, J 
Phys C 18, L 927 (1985).

\bibitem{dsfisher} D S Fisher, Phys Rev Lett 56, 416 (1986).

\bibitem{sethna3} J P Sethna, K A Dahmen, and C R Myers, Nature (London) 
410, 242 (2001) and references therein.

\bibitem{dhar} D Dhar, P Shukla, J P Sethna, J Phys A30, 5259 (1997).

\bibitem{shukla} P Shukla, Prog Theo Phys, Vol 96, No 1 (1996); P 
Shukla, Physica A 233,235 (1996); P Shukla,Phys Rev E 62, 4725 (2000); 
P Shukla,Phys Rev E 63,027102 (2001).

\bibitem{shukla2} P Shukla and R S Kharwanlang, Phys Rev E 81, 031106 
(2010); P Shukla and R S Kharwanlang, Phys Rev E 83, 011121 (2011).

\bibitem{dunlop1} {\em{Rock Magnetism: Fundamentals and Frontiers}} by D 
J Dunlop and O Ozdemir, Cambridge University Press, 2001.

\bibitem{dunlop2} D J Dunlop. Physics of the Earth and Planetary 
Interiors 26, 1-26 (1981).

\bibitem{note} An important aspect of rock magnetism (not discussed 
here) is that it provides a recorded memory of the intensity, direction, 
and polarity of the Earth's magnetic field in the geological past. It 
reveals that the polarity of Earth's magnetic field has reversed several 
times in the past, most recently about 780,000 years ago. This discovery 
revolutionized the field in the 1960s by providing evidence for 
continental drift and plate tectonics. See for example, D J Dunlop, 
"Rock magnetism" in AccessScience, @McGraw-Hill Companies, 
2008,http://www.accessscience.com

\bibitem{roberts1} A P Roberts, Y Cui, K L Verosub. J Geo Phys Res 100, 
17909-17924 (1995), and references therein.

\bibitem{roberts2} A P Roberts, C R Pike, K L Verosub. 
J Geo Phys Res 105, 28461-28475 (2000).

\bibitem{tauxe1} L Tauxe, T A T Mullender, T Pick. J Geo Phys Res 101, 
571-583 (1996).

\bibitem{wasilewski} P J Wasilewski, Earth Planet Sci Lett 20, 67-72 
(1973).

\bibitem{becker} J J Becker. IEEE Trans Magn Mag 18,1451-1453 (1982).

\bibitem{pick1} T Pick, L Tauxe. J Geo Phys Res 98,17949-17964 (1993).

\bibitem{pick2} T Pick, L Tauxe. Geo Phys J Int 119, 116-128 (1994); 
E502 (2005).

\bibitem{bean} C P Bean, J Appl Phys 26,11 (1955).

\bibitem{hejda} P Hejda, A Kapicka, E Petrovsky, B A Sjoberg. IEEE Trans 
Magn 30,2 (1994).

\bibitem{bennett} L H Bennett and E D Torre. J Appl Phys 97 10E502 
(2005).

\bibitem{davies} J E Davies, L H Bennett, E D Torre, B C Choi, S N 
Piramanayagam, and E Girgis, IEEE Transactions on Magnetics, vol 44, 
2722-2725 (2008).

\bibitem{tauxe2} L Tauxe, H N Bertram, and C Seberino, Geochem. Geophys. 
Geosyst., 3 (10), 1055 (2002). doi:10.1029/2001GC000241

\bibitem{kletetschka} G Kletetschka, and P J Wasilewski, Physics of the 
Earth and Planetary Interiors 129, 173-179 (2002).

\bibitem{liu} Y Liu and K A Dahmen, Phys Rev E 79, 061124 (2009).

\bibitem{dahmen} K Dahmen and J P Sethna, Phys Rev B 53, 14872 (1996).

\bibitem{straka} L Straka, O Heczko, and N Lanska, IEEE Trans Magn 38, 5 
(2002)

\bibitem{murray} S J Murray, M Marioni, S M Allen, R C O'Handley, and T 
A Lograsso, Appll Phys Lett 77, 6 (2000) 

\bibitem{goicoechea} J Goicoechea and J Ortin, J Phys IV 
France 05, C2-71 (1995).


\end{thebibliography}
\end{document}